\documentclass[aps,prd,tightenlines,nofootinbib]{revtex4}

\usepackage{graphicx}
\usepackage{dcolumn}
\usepackage{bm}
{\def\mtiny{\vrule width 0pt}
\def\mrm#1{\mathrm{#1}}
\def\DZ{\relax\ifmmode{D^0}\else{$\mrm{D}^{\mrm{0}}$}\fi}
\def\DZB{\relax\ifmmode{\overline{D}\mtiny^0}\else{$\overline{\mrm{D}}\mtiny^{\mrm{0}}$}\fi}

\begin{document}

{\def\mtiny{\vrule width 0pt}
\def\DZ{\relax\ifmmode{D^0}\else{$\mrm{D}^{\mrm{0}}$}\fi}
\def\DZB{\relax\ifmmode{\overline{D}\mtiny^0}\else{$\overline{\mrm{D}}\mtiny^{\mrm{0}}$}\fi}
}
\begin{flushright} 
CLNS 03/1848\\       
CLEO 03-13\hphantom{48}
\end{flushright}         

\title{Search for $CP$ Violation in $\DZ \to K_S^0\pi^+\pi^-$}  


\author{D.~M.~Asner}
\author{H.~N.~Nelson}
\affiliation{University of California, Santa Barbara, California 93106}
\author{R.~A.~Briere}
\author{G.~P.~Chen}
\author{T.~Ferguson}
\author{G.~Tatishvili}
\author{H.~Vogel}
\affiliation{Carnegie Mellon University, Pittsburgh, Pennsylvania 15213}
% ...
\author{(CLEO Collaboration)} 
\noaffiliation


\date{November 12, 2003}

\begin{abstract} 
We report on a search for $CP$ violation in the decay of $\DZ$ and $\DZB$ to $K_S^0\pi^+\pi^-$. The data come from an integrated luminosity of $9.0$~fb$^{-1}$ of $e^+e^-$ collisions at ${\sqrt s}\approx 10$~GeV recorded with
the CLEO~II.V detector. The resonance substructure of this decay is well
described by ten quasi-two-body decay channels 
plus a small non-resonant component.
We observe no evidence for $CP$ violation in the amplitudes that
describe the decay $\DZ \to K_S^0\pi^+\pi^-.$  
\end{abstract}

\pacs{13.25.Ft,11.30.Er}
\maketitle

Phenomena that are not invariant with respect to charge conjugation
and parity ($CP$)
in strange~\cite{Alavi-Harati:1999xp,Fanti:1999nm} 
and bottom~\cite{Aubert:2001nu,Abe:2002px}
mesons are the motivation for numerous current and future experiments.
Standard Model (SM) predictions for the rate of $CP$ violation in charm mesons are as large
as 0.1\% for $\DZ \to
\pi^+\pi^-\pi^0$~\cite{Buccella:1996uy,Santorelli:1996bc}
but are considerably
smaller, ${\cal O} (10^{-6})$, for $\DZ \to K^0_S \pi^+\pi^-$ where
the
dominant contribution is due to $K^0-\overline K^0$ mixing~\cite{Xing:1995jg}.
The Dalitz 
technique~\cite{dalitz,bergfeld} 
allows increased sensitivity to $CP$ violation by probing the 
decay amplitude rather than the decay rate. Observation of $CP$ violation in $\DZ \to K^0_S \pi^+\pi^-$ at current experimental 
sensitivity would be strong evidence
for non-SM processes.
The decay $B^\pm \to DK^\pm$ followed by a multibody $\DZ$ decay, such as 
$\DZ \to K^0_S \pi^+\pi^-$, may elucidate the origin of $CP$ violation in the 
$B$ sector~\cite{Giri:2003ty}.

We present the results of a search for $CP$ violation in the 
amplitudes that contribute to $\DZ \to K^0_S \pi^+\pi^-$. 
Previous searches for direct $CP$ violation~\cite{kpi,Frabetti:1994kv,Bartelt:1995vr,Aitala:1997ff,Link:2000aw} in 
the neutral charm meson system set limits of a few percent.



This analysis uses an integrated luminosity of 9.0~fb$^{-1}$
of $e^+e^-$ collisions at $\sqrt{s}\approx10\,$GeV provided by
the Cornell Electron-positron Storage Ring (CESR).
The data were taken with the CLEO~II.V 
detector~\cite{ctwo}. 

The event selection is identical to that used in our previous analysis of
$\DZ \to K^0_S \pi^+\pi^-$~\cite{ourprl} which did not consider $CP$ violation.
We reconstruct candidates for the decay sequence
$D^{\ast+}\!\to\!\pi^+_S \DZ$, $\DZ\!\to\!K^0_S\pi^+\pi^-$.
The charge of the slow
pion ($\pi^+_S$ or $\pi^-_S$) identifies the charm state 
as either $\DZ$ or $\DZB$. 
Consideration of charge conjugation is implied
throughout this paper, unless otherwise stated.

We evaluate the energy released in the
$D^{\ast+}\!\to\!\pi^+_S\DZ$ decay as
$Q\!\equiv\!M^\ast\!-\!M\!-\!m_\pi$,
where $M^\ast$ is the reconstructed mass of the
$\pi_S^+ K^0_S \pi^+ \pi^-$ system, $M$ is the reconstructed mass of
the
$K^0_S \pi^+\pi^-$ system, and $m_\pi$ is the charged pion mass.
\begin{figure}
\includegraphics*[width=10.25in]{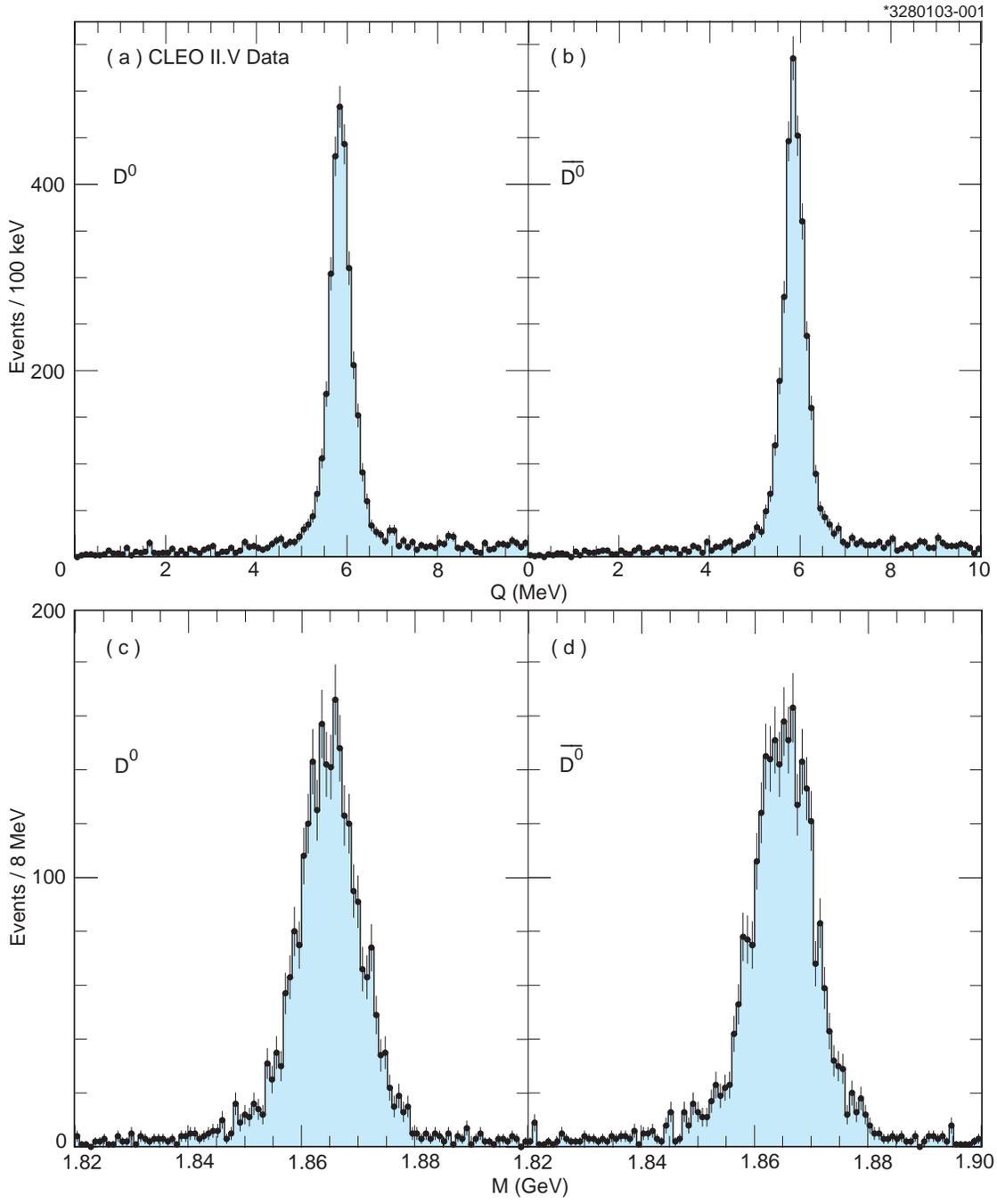}
\vskip -5.cm
\caption{Distribution of $Q$, (a) and (b), and $M$, (c) and (d),
for the process $\DZ$ and $\DZB\to K^0_S\pi^+\pi^-$.
The candidates  
pass all selection criteria discussed in the text.}
\label{fig:qandm}
\end{figure}
We require the $D^{\ast+}$ momentum
$p_{D^\ast}$ to exceed 2.0 GeV$/c$. We reconstruct $K^0_S\!\to\!\pi^+\pi^-$ with the
requirement that the daughter pion tracks form a common vertex, 
in three dimensions, with a confidence level $>10^{-6}$.
Signal candidates pass
the vertex requirement with $96\%$ relative efficiency. Throughout this paper,
relative efficiency is defined as the number of events in the data
passing all 
requirements relative to the number of events when only the
requirement under
study is relaxed.

Our silicon vertex detector provides precise measurement of charged 
tracks in three dimensions~\cite{svxres}.
We exploit the precision tracking by refitting the 
$K_S^0$ trajectory and $\pi^\pm$ tracks with a requirement that they
form a common vertex in three dimensions.
We use the trajectory of the
$K^0_S\pi^+\pi^-$ system and the position of the CESR luminous region
to obtain the $\DZ$ production point.  
We refit the $\pi_S^+$ track with a requirement that
the trajectory intersect
the $\DZ$ production point.
We require the
confidence level of each refit exceed $10^{-4}$. 
The signal candidates pass the $\DZ$ production and decay vertex requirement 
with $85\%$ and $91\%$ relative efficiency, respectively.

We select 5,299 candidates 
within three standard deviations of
the expected $Q$,
$M$, and $m_{K_S^0}$. 
We compute $\sigma_Q$, $\sigma_M$ and $\sigma_{m_{K_s^0}}$
from the trajectory reconstruction covariance matrices of the
daughters of each $D^{\ast+}$ candidate.
The distributions of $Q$ and $M$
for the $\DZ$ and $\DZB$
samples
for our data are shown in Fig.~\ref{fig:qandm}. 
We find 2,579 $\DZ$ and 2,720 $\DZB$ candidates
corresponding to an asymmetry of $(-2.7\pm 1.4 \pm 0.8)\%$, a
1.7$\sigma$ effect.

\begin{figure} 
\includegraphics*[width=6.25in]{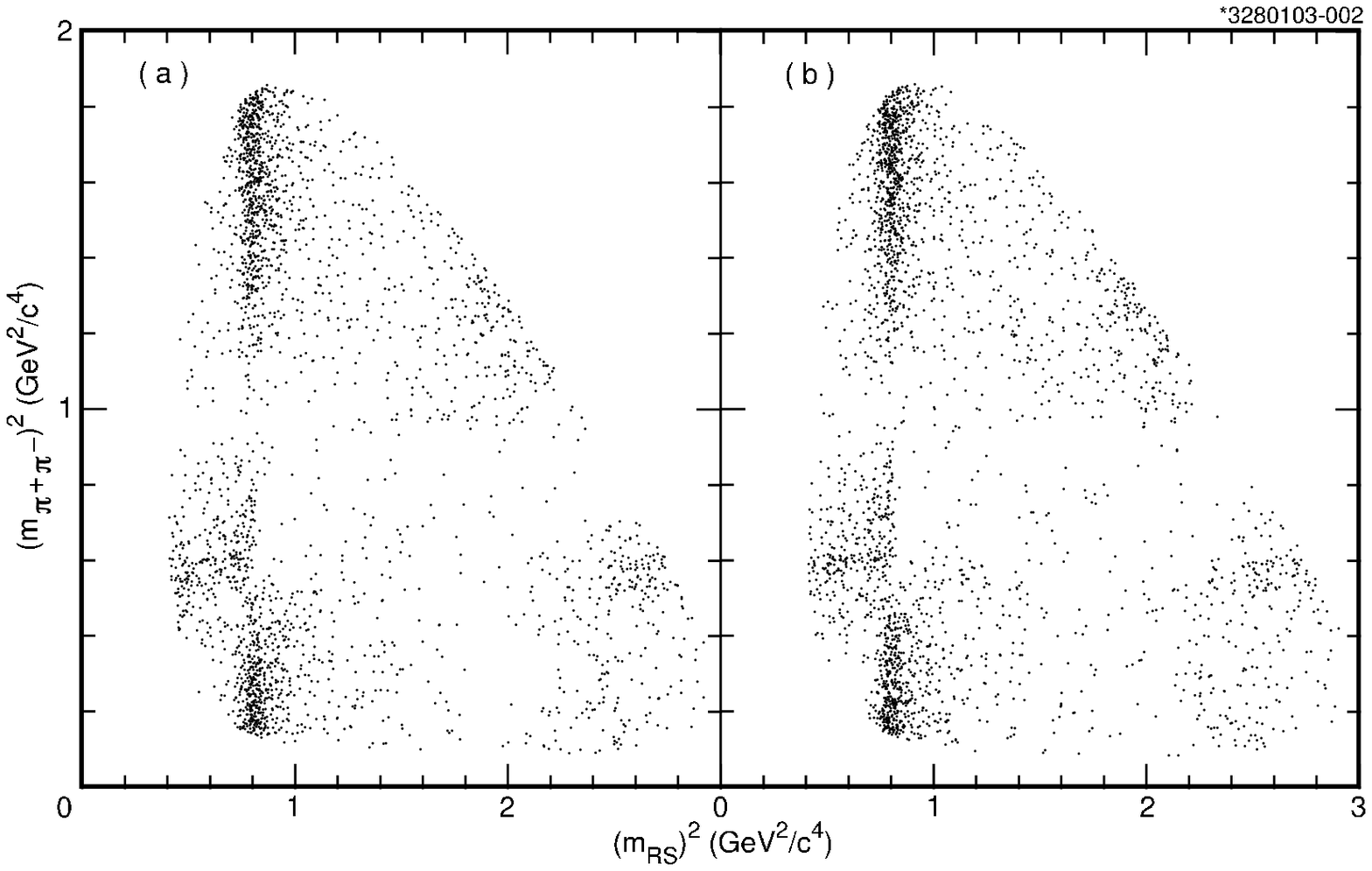}
\caption
{\label{fig:dal}
Dalitz distribution for a) $\DZ\to K^0_S\pi^+\pi^-$ and b)
$\DZB\to K^0_S\pi^+\pi^-$ candidates  
passing all selection criteria discussed 
in Ref.~[17].
The horizontal
axis $(m_{RS})^2$ corresponds to $(m_{K^0_S\pi^-})^2$ for $\DZ$ and 
$(m_{K^0_S\pi^+})^2$ for $\DZB$.}
\end{figure}

In Fig.~\ref{fig:dal}, we plot $m^2_{RS}$ vs $m^2_{\pi^+\pi^-}$ where $m_{RS}$ 
denotes the `right-sign' and corresponds to $m_{K^0_S\pi^-}$ for $\DZ$ and
$m_{K^0_S\pi^+}$ for $\DZB$.
Similarly, $m_{WS}$ denotes  the `wrong-sign' and corresponds to $m_{K^0_S\pi^+}$ for $\DZ$
and $m_{K^0_S\pi^-}$ for $\DZB$.
We study our efficiency with a GEANT~\cite{GEANT} based simulation of
$e^+ e^- \to c \overline c$ events in our detector with a luminosity corresponding
to more than three times our data sample.  
We observe that our selection introduces distortions due
to inefficiencies near the edge of phase space, and fit the efficiency to a two
dimensional cubic
polynomial ${\cal E}(m_{RS}^2$, $m_{\pi^+\pi^-}^2)$.
The reconstruction efficiencies for the $\DZ$ and $\DZB$ 
over the Dalitz plot are consistent with each other and we take them to 
be equal; so \hbox{${\cal E}(m_{RS}^2$, $m_{\pi^+\pi^-}^2) = {\cal E}(m_{WS}^2$, $m_{\pi^+\pi^-}^2) = {\cal E}$}.

Figure~\ref{fig:qandm} shows that
the background is small, but non-negligible, and we model our
background as in Ref.~\cite{ourprl}.
To model the background contribution in the Dalitz distribution we consider 
those
events in the data and MC that are in sidebands five to ten standard deviations
from the signal in $Q$ and $M$ and within three in $m_{K^0_S}$. 
There are 235 (579) $\DZ$ and 210 (572) $\DZB$  candidates in this
selection in the data (MC), about four times the amount
of background we estimate from the signal region.  We constrain the
invariant mass of the sideband candidates to the $\DZ$ mass and compare 
with the background in our signal region 
from our simulation which also includes $e^+e^-$ annihilations producing
the lighter quarks. We note that the background from the simulation is 
dominated by random combinations of unrelated tracks.
The simulation predicts that the background uniformly populates
the allowed phase space.
We model this contribution to the Dalitz distribution by fitting the
$\DZ$ mass constrained data sideband sample to a two dimensional cubic
polynomial ${\cal B}(m_{RS}^2$,$m_{\pi^+\pi^-}^2)$. All parameters
except the constant are consistent with zero as predicted by simulation,
so ${\cal B}(m_{RS}^2$,$m_{\pi^+\pi^-}^2)={\cal
B}(m_{WS}^2$,$m_{\pi^+\pi^-}^2)={\cal B}$. 
The normalization of the uniform background in the data exceeds the
simulation by $21\pm 8\%$.
Other possible contributions to the background, where 
$\pi\pi$ or $K\pi$ resonances
combined with random tracks fake a $\DZ$ and real $\DZ$ decays combined with random
soft pions of the wrong charge.
The latter,
called mistags,
are especially dangerous to our search for $CP$ violation as $\DZ$
candidates are misidentified as $\DZB$'s.
Since only two of the
Dalitz parameters are independent~\cite{pdg},
mistags populate the Dalitz distribution in a known way that depends on the
shape of signal - namely we interchange assignment of $m_{RS}$ and $m_{WS}$.  
When we analyze the Dalitz distribution, we allow a mistag
fraction with an unconstrained contribution 
and 
we have looked for the
contribution of a resonance, such as $\rho^0$ or
$K^\ast(892)^-$, plus random tracks to the background in the data
and conclude that any such contributions are negligible.

We parameterize the $K^0_S\pi^+\pi^-$ Dalitz distribution using 
the isobar model described in Ref.~\cite{bergfeld} where each resonance, $j$, 
has its own amplitude, $a_j$, and phase, $\delta_j$. 
A second process, not necessarily of SM origin, 
could contribute to the $j$-th resonance. 
%
In general, we can express the amplitudes to the $j$-th quasi-two-body state as
$(a_j e^{i(\delta_j\pm\phi_j)}\!\pm\! b_j e^{i(\delta_j\pm\phi_j)}){\cal A}_j\! =\! a_j e^{i(\delta_j\pm\phi_j)}(1\! \pm\! \frac{b_j}{a_j}){\cal A}_j,$
with `$+$' for $\DZ$ and `$-$' for $\DZB$ and
${\cal A}_j = {\cal A}_j(m_{RS}^2,m_{\pi\pi}^2)$ is the amplitude for resonance $j$ as
described in Ref.~\cite{bergfeld}. Thus $a_j$ and
$\delta_j$ are explicitly $CP$ conserving amplitudes and phases,
$b_j$ are explicity $CP$ violating amplitudes normalized by the
corresponding $CP$
conserving amplitude $a_j$, and $\phi_j$ are explicitly $CP$
violating phases. In the absence
of $CP$ violation $b_j$ and $\phi_j$ would be zero.
The matrix elements ${\cal M}$ and ${\cal \overline M}$ for the $\DZ$ and $\DZB$ samples, respectively, 
are defined as
\begin{eqnarray}
{\cal M} & = & a_0e^{i\delta_0} + \sum_j a_j  e^{i(\delta_j+\phi_j)}(1 + \frac{b_j}{a_j}){\cal A}_j \\
{\cal \overline M} & = & a_0e^{i\delta_0} + \sum_j a_j  e^{i(\delta_j-\phi_j)}(1 - \frac{b_j}{a_j}){\cal A}_j, 
\end{eqnarray}
where $a_0$ and $\delta_0$ parametrize the non-resonant amplitude, assumed to
be $CP$ conserving.

We perform an unbinned maximum likelihood fit which minimizes the
function
\begin{equation}
{\cal F}\!=\!{\big[\sum_{\DZ}-2\ln {\cal L}\big]\!+\!\big[\sum_{{\DZB}}-2\ln {\cal \overline L}\big]\!+\!\left( {F\!-\!F_o \over{\sigma_F}} \right)^2}
\end{equation}
where
\begin{eqnarray}
\label{eq:likelihood}
& {\cal L} = 
T\left(F{{{\cal E}
\left| {\cal M}(m^2_{RS}, m^2_{\pi\pi}) \right|^2 }\over{ {\cal N}}} 
 \!+\! (1\!-\!F) {{{\cal B}}\over{{\cal N}_{\rm background}}}\right) \\ \nonumber 
& \!+\! (1\!-\!T) \left(F{{{\cal E}
\left| {\cal M}(m^2_{WS}, m^2_{\pi\pi}) \right|^2 }\over{ {\cal N}}}
 \!+\! (1\!-\!F) {{{\cal B}}\over{{\cal N}_{\rm background}}}\right) 
 \\
& {\cal \overline L} =  T\left(F{{{\cal E}
\left| {\cal \overline M}(m^2_{RS}, m^2_{\pi\pi}) \right|^2 }\over{ {\cal N}}}
\!+\! (1\!-\!F) {{{\cal B}}\over{{\cal N}_{\rm background}}}\right)
\\ \nonumber & \!+\! (1\!-\!T)
\left(F{{{\cal E}
\left| {\cal \overline M}(m^2_{WS}, m^2_{\pi\pi}) \right|^2 }\over{ {\cal N}}}
\!+\! (1\!-\!F) {{{\cal B}}\over{{\cal N}_{\rm background}}}\right)
\end{eqnarray}
%
and
%
\begin{eqnarray}
& {\cal N} \equiv \frac{1}{2}\left(\int {{\cal E}
 \left| {\cal M} \right|^2} dm^2_{RS}dm^2_{\pi\pi}\!+\!\int {{\cal E} \left| {\cal \overline M} \right|^2} dm^2_{RS}dm^2_{\pi\pi}\right) 
\label{eqn:normm} \\
%
& {\cal N}_{\rm background} \equiv \int {\cal B} dm^2_{RS}dm^2_{\pi\pi}
\label{eqn:normbkgd}
\end{eqnarray}
define the normalization of the $\DZ$, $\DZB$, and background events.
The signal fraction $F_o$ and its error $\sigma_F$
are determined from the fit to the combined $\DZ$ and $\DZB$ ~mass spectra, shown in Fig.~\ref{fig:qandm}c) and d), to be $0.979\pm0.015$.
The signal fraction $F$ and the mis-tag fraction $(1-T)$ 
are consistent in the $\DZ$ and $\DZB$ samples and we take them to be equal.
The signal fraction $F$, 
the mistag fraction $(1-T)$, and the parameters $a_j, \delta_j, b_j,$ and $\phi_j$ that
describe the matrix elements ${\cal M}$ and ${\cal \overline M}$ are determined by the 
Dalitz plot fit.
We test the performance of our fit by generating 100 Monte Carlo samples 
from the results of our standard fit in Ref.~\cite{ourprl}. The pull distributions of all fit parameters are consistent with unit Gaussian with zero mean
indicating that the fit is not biased and that the errors are 
correctly computed.
%
%

We begin our search for $CP$ violation from the results of our standard
fit in Ref.~\cite{ourprl} which clearly observed the ten modes,
($K^{\ast-} \pi^+$,
${K}^{\ast}_{0}\!(1430)^- \pi^+$,
${K}^{\ast}_{2}\!(1430)^- \pi^+$,
${K}^{\ast}(1680)^- \pi^+$,
$K^0_S \rho$,
$K^0_S \omega$,
$K^0_S f_0(980)$,
$K^0_S f_2(1270)$,
$K^0_S f_0(1370)$,
and the ``wrong sign'' $K^{\ast+} \pi^-$) plus a small non-resonant component.
First, we fit the $\DZ$ and $\DZB$ samples independently. 
The results of the $\DZ$ and $\DZB$ fits are consistent with 
each other and with our $CP$ conserving result~\cite{ourprl}.
Next, we fit the $\DZ$ and $\DZB$ samples simultaneously.
This fit has 42 free parameters, ten $CP$ conserving amplitudes and ten $CP$ conserving phases which are the same in the $\DZ$ and $\DZB$ samples,
ten $CP$ violating amplitudes and ten $CP$ violating phases which differ by a
sign in the $\DZ$ and $\DZB$ samples, plus 
two normalizations for the combinatoric and mistag backgrounds.

\begin{figure}
\includegraphics*[width=6.25in]{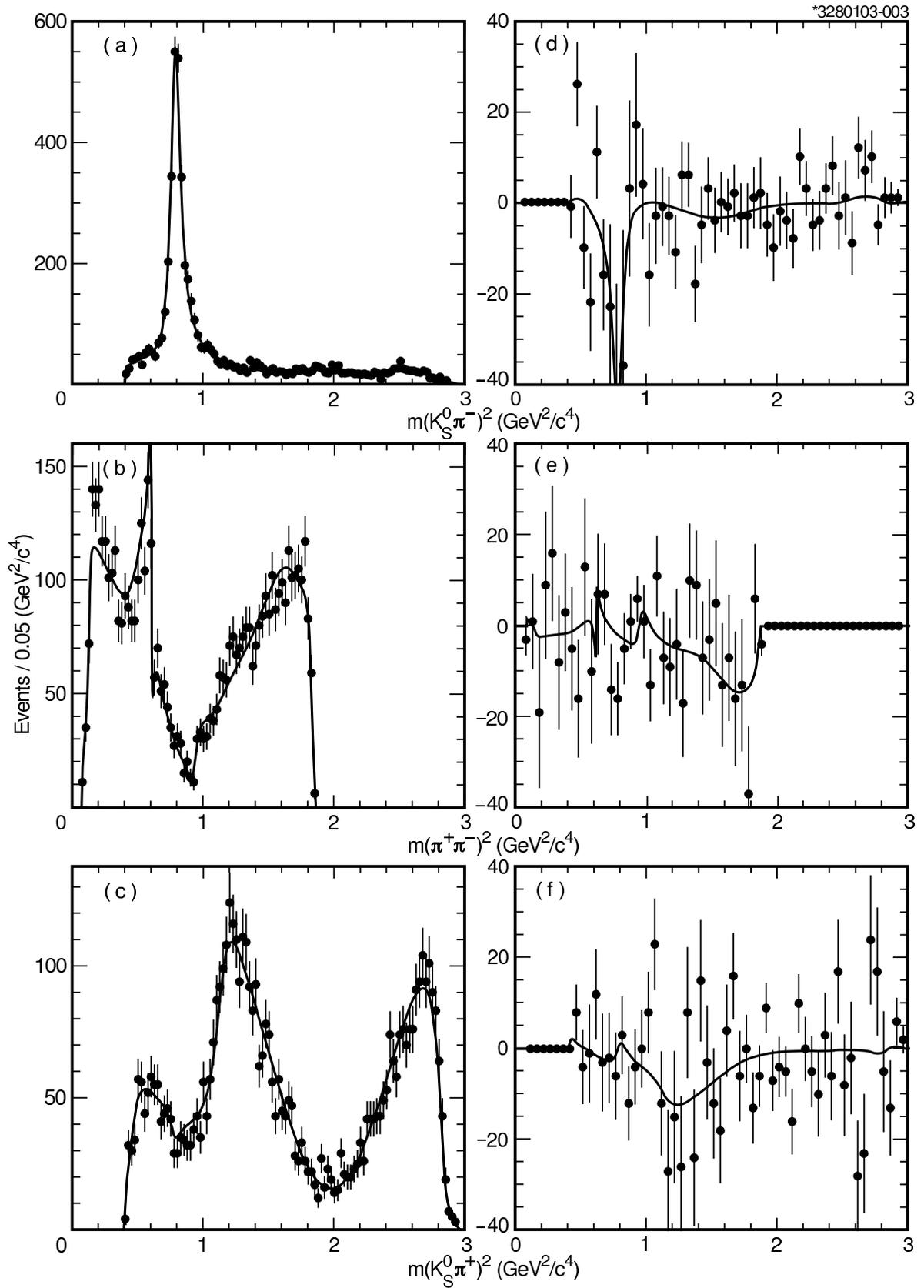}
\caption
{\label{fig:projsum}
Projections of the results of the fit described in the text
to the $K^0_S\pi^+\pi^-$~Dalitz distribution showing both the
fit (histogram) and the data (points).  The combined $\DZ$ and $\DZB$ samples are shown in (a), (b) and (c).  The difference between the
$\DZ$ and $\DZB$ samples is shown in (d), (e) and (f).}
\end{figure}

We report the $CP$ conserving amplitude and phase, $a_j$ and $\delta_j$, 
in Table~\ref{tbl:fit1}. These results are consistent with 
our result in Ref.~\cite{ourprl} in which $CP$
conservation was assumed. We report
the fractional $CP$ violating amplitude and $CP$ violating phase,
$b_j/a_j$ and $\phi_j$ in Table~\ref{tbl:fit2}. 
The three projections of the fit to the combined $\DZ$ and $\overline
\DZ$ samples
and the difference between the $\DZ$ and $\DZB$ samples are shown
in Fig.~\ref{fig:projsum}. 
We find the signal fraction $F$ and mistag fraction $(1-T)$ to be, $97.1\pm0.8\%$ and 
$0.0^{+0.7}_{-0.0}\%$, respectively.
The confidence level of the fit, calculated directly from the
likelihood function~\cite{argus, bergfeld}, is 55\%. 

The amplitude $a_j$, on its own, is not a good estimator of the
contribution of resonance $a_j$ to the total rate. The width of a
resonance, interference with other resonances and the allowed phase
must be considered.
The fit fraction (FF), formulated to encapsulate all these effects,
is commonly defined as the
integral of a single amplitude squared
over the Dalitz plot
($m^2_{RS}$ vs $m^2_{\pi\pi}$) divided by the coherent sum of all amplitudes
squared. We define the fit fraction as
\begin{equation}
\label{eqn:ddpintD0}
{\rm FF}_j \! =  \! \frac{\int \! \left|(a_j + b_j){\cal A}_j \right|^2\!dm^2_{RS}dm^2_{\pi\pi}} 
{\int \left|{\cal M}\right|^2 dm^2_{RS}dm^2_{\pi\pi}}
\end{equation}
\begin{equation}
 \label{eqn:ddpintD0B}
{\rm \overline{ FF}}_j \! =  \!
\frac{\int \! \left|(a_j  - b_j){\cal A}_j \right| ^2\!dm^2_{RS}dm^2_{\pi\pi}} 
{\int \left|{\cal \overline M}\right|^2 dm^2_{RS}dm^2_{\pi\pi}}
\end{equation}
for the $\DZ$ and $\DZB$ samples, respectively. The $CP$ conserving and $CP$
violating fit fractions are defined as the sum and difference of the
numerators of Eq.~\ref{eqn:ddpintD0} and Eq.~\ref{eqn:ddpintD0B},
respectively, divided by the sum of the denominators of Eq.~\ref{eqn:ddpintD0} and Eq.~\ref{eqn:ddpintD0B}.

The dominant constraint on $CP$ violation is not due to limits on
the $CP$ violating amplitude squared but is due to the potential interference of a $CP$
violating amplitude with a well determined $CP$ conserving amplitude.
We define the $CP$ violating interference fraction (IF) as 
\begin{equation}
\label{eqn:ifddpint}
{\rm IF}_j \!=\!  {\left|\int
\sum_k \left( 2a_k e^{i\delta_k}\sin(\phi_k+\phi_j){\cal A}_k\right) b_j 
{\cal A}_j dm^2_{RS}dm^2_{\pi\pi}\right| \over
\left({\int \left|{\cal M}\right|^2
dm^2_{RS}dm^2_{\pi\pi} +
\int \left|{\cal \overline M}\right|^2
dm^2_{RS}dm^2_{\pi\pi}}\right)}
\end{equation}
where for the non-resonant component ${\cal A}_0=1$.
The value of $b_j$ determined by our fit is constrained by terms in the 
likelihood function proportional to $|b_j|^2$ and $a_ke^{+i\delta_k}b_j$ which are sensitive
to both $CP$ violation in the direct decay to a given
submode and possible $CP$ violation in interference with other modes, respectively.
The $CP$ violating fit fraction defined by Eqs.~\ref{eqn:ddpintD0} and~\ref{eqn:ddpintD0B} is
sensitive to $CP$ violation in decay.
The $CP$ violating interference fractions of Eq.~\ref{eqn:ifddpint} sum over
the contribution proportional to $a_ke^{+i\delta_k}b_j$ so are sensitive to $CP$ violation in interference. The phases are
important and allow the possibility of cancellation in this sum.
This makes the IF
a better representation of the impact of $CP$ violation on the rate of decay
than the $CP$ violating FF.
To quantify a fractional $CP$ violation each decay channel, we define
\begin{equation}
\label{eqn:acpj}
A_{CP_j} = {\rm IF}_j/{\rm FF}_j,
\end{equation}
where ${\rm FF}_j$ are the $CP$ conserving fit fractions.

Note that the observation of $CP$ violation with a significantly non-zero
$b_j/a_j$ or $\phi_j$ does not imply
that \emph{any} of the derived quantities, the $CP$ violating FF$_j$, IF$_j$ or 
$A_{CP_j}$ will be non-zero. In particular, the $CP$ violating effects that are locally non-zero can integrate to zero, thus IF$_j$ and $A_{CP_j}$ are more sensitive measures of $CP$ violation than the $CP$ violating fit fraction.



\begin{table*}

\caption{
$CP$ Conserving Parameters. Errors are
statistical, experimental systematic and modeling systematic,
respectively. The $CP$ conserving 
fit fraction is computed from Eqs.~\ref{eqn:ddpintD0} and~\ref{eqn:ddpintD0B}
following the prescription described in the text.}
\label{tbl:fit1}
\begin{ruledtabular}
\begin{tabular}{l|ccc}
Component     & Amplitude ($a_j$) & 
Phase ($\delta_j$) ($^\circ$) &
Fit Fraction (\%) \\ \hline
                                        $K^\ast(892)^+\pi^- \times B(K^\ast(892)^+\to K^0 \pi^+)$                        & $(11\!\pm\! 2\!^{+ 4}_{-1}\!^{+ 2}_{- 1})\times 10^{-2}$ & $  324\!\pm\!  12\!^{+  9}_{-  2}\!^{+  6}_{- 13}$ & $ 0.34\!\pm\! 0.13\!^{+ 0.35}_{- 0.04}\!^{+ 0.06}_{- 0.03}$                                                                                                     \\
                                        $\overline K^0 \rho^0$                                                          & $ 1.0$ (fixed) & $0$ (fixed) & $26.7\!\pm\!1.1\!^{+ 0.7}_{- 0.8}\!^{+ 0.5}_{- 2.7}$                                                                                                     \\
                                        $\overline K^0 \omega \times B(\omega \to \pi^+\pi^-)$                          & $(40\!\pm\! 5\!\pm\!2\!^{+ 7}_{- 1})\times 10^{-3}$ & $115\!^{+  6}_{-  7}\!^{+  4}_{-  2}\!^{+  3}_{-  2}$ & $ 0.81\!\pm\! 0.19\!^{+ 0.07}_{- 0.08}\!^{+ 0.17}_{- 0.06}$                                                                                                     \\
                                        $K^\ast(892)^-\pi^+ \times B(K^\ast(892)^-\to{\overline K}^0\pi^-)$              & $ 1.54\!\pm\! 0.04\!^{+ 0.01}_{- 0.02}\!^{+ 0.11}_{- 0.02}$ & $150\!\pm\!  2\!\pm\!2\!\pm\!2$ & $66.3\!\pm\! 1.3\!^{+ 0.7}_{- 2.7}\!^{+ 2.3}_{- 3.3}$                                                                                                     \\
                                        $\overline K^0 f_0(980) \times B(f_0(980) \to \pi^+ \pi^-)$                     & $ 0.34\!\pm\! 0.02\!^{+ 0.05}_{- 0.02}\!\pm\!0.01$ & $188\!\pm\!  5\!^{+  7}_{-  4}\!^{+  9}_{-  4}$ & $ 4.2\!\pm\! 0.5\!^{+ 1.1}_{- 0.4}\!\pm\!0.2$                                                                                                     \\
                                        $\overline K^0 f_2(1270) \times B(f_2(1270) \to \pi^+ \pi^-)$                   & $ 0.79\!\pm\! 0.23\!^{+ 0.27}_{- 0.13}\!^{+ 0.39}_{- 0.51}$ & $306\!^{+  13}_{- 15}\!^{+  7}_{- 23}\!^{+ 54}_{-  3}$ & $ 0.36\!\pm\! 0.22\!^{+ 0.31}_{- 0.12}\!^{+ 0.08}_{- 0.15}$                                                                                                     \\
                                        $\overline K^0 f_0(1370) \times B(f_0(1370) \to \pi^+ \pi^-)$                   & $ 1.74\!\pm\! 0.13\!^{+ 0.23}_{- 0.22}\!^{+ 0.10}_{- 0.20}$ & $ 85\!\pm\!  5\!^{+  5}_{-  1}\!^{+  2}_{- 12}$ & $ 9.8\!\pm\! 1.4\!^{+ 2.4}_{- 2.1}\!^{+ 1.1}_{- 2.9}$                                                                                                     \\
                                        $K^\ast_0(1430)^-\pi^+ \times
                                        B(K^\ast_0(1430)^-\to\overline
                                        K^0\pi^-)$         & $ 1.93\!\pm\!0.11\!^{+ 0.04}_{- 0.17}\!^{+ 0.30}_{- 0.09}$ & $  1\!\pm\! 5\!^{+  6}_{-  4}\!^{+  8}_{- 14}$ & $ 7.2\!\pm\!0.7\!^{+ 0.3}_{- 1.1}\!^{+ 1.4}_{- 0.7}$                                                                                                     \\
                                        $K^\ast_2(1430)^-\pi^+ \times
                                        B(K^\ast_2(1430)^-\to\overline
                                        K^0\pi^-)$         & $ 0.94\!^{+ 0.12}_{- 0.11}\!^{+ 0.13}_{- 0.10}\!^{+ 0.00}_{- 0.06}$ & $335\!^{+ 9}_{-  8}\!^{+ 0}_{-  6}\!^{+  4}_{- 23}$ & $ 1.1\!\pm\! 0.2\!^{+ 0.3}_{- 0.2}\!^{+ 0.4}_{- 0.2}$                                                                                                     \\
                                        $K^\ast(1680)^- \pi^+ \times
                                        B(K^\ast(1680)^- \to \overline
                                        K^0 \pi^-)$         & $5.49\!^{+ 0.67}_{- 0.65}\!^{+ 0.74}_{- 0.94}\!\pm\!3.9$ & $ 175\!\pm\! 7\!^{+ 11}_{-  6}\!^{+ 16}_{-  3}$ & $ 2.3\!\pm\!  0.5\!^{+ 0.6}_{- 0.7}\!^{+ 0.3}_{- 1.2}$                                                                                                     \\
                                        $\overline K^0 \pi^+ \pi^-$ non-resonant                                        & $ 0.93\!^{+ 0.34}_{- 0.31}\!^{+ 0.59}_{- 0.17}\!^{+ 0.92}_{- 0.53}$ & $343\!^{+ 23}_{- 16}\!^{+ 58}_{- 23}\!^{+ 74}_{- 17}$ & $ 0.7\!\pm\!0.7\!^{+ 1.1}_{- 0.2}\!^{+ 1.8}_{- 0.6}$                                                                                                     \\

\end{tabular} 
\end{ruledtabular}

\caption{
$CP$ Violating Parameters. Errors are
statistical, experimental systematic and modeling systematic,
respectively. The interference fraction, $CP$ violating fit fraction and 
$A_{CP}$, computed from Eq.~\ref{eqn:ddpintD0}-Eq.~\ref{eqn:acpj},
following the prescription described in the text, include
statistical and systematic effects.}
\label{tbl:fit2}
\begin{ruledtabular}
\begin{tabular}{l|ccccc}
Component     & Ratio ($b_j/a_j$) & 
Phase ($\phi_j$) ($^\circ$) &
Interference & $CP$ Violating & $A_{CP}(\%)$ \\
& & & Fraction & Fit Fraction   & \\
& & & \multicolumn{3}{c}{($95\%$ Upper Limits)} \\ \hline
                                        $K^\ast(892)^+\pi^- \times B(K^\ast(892)^+\to K^0 \pi^+)$                        & $-0.12\!^{+0.21}_{-0.22}\!^{+0.09}_{-0.15}\!^{+0.11}_{-0.03}$ & $  6\!^{+ 21}_{- 22}\!^{+ 13}_{- 35}\!^{+ 18}_{-  4}$ & $< 3.0\times 10^{-3}$ & $< 7.8\times 10^{-4}$ & $<92$                                                                                                  \\
                                        $\overline K^0 \rho^0$                                                          & $0.001\!\pm\!0.022\!^{+0.011}_{-0.009}\!^{+0.002}_{-0.011}$ & $ -1\!^{+ 16}_{- 18}\!^{+  9}_{- 31}\!^{+ 21}_{-  3}$ & $< 0.7\times 10^{-3}$ & $< 4.8\times 10^{-4}$ & $< 0.3$                                                                                                  \\
                                        $\overline K^0 \omega \times B(\omega \to \pi^+\pi^-)$                          & $-0.14\!^{+0.10}_{-0.11}\!^{+0.11}_{-0.01}\!^{+0.01}_{-0.02}$ & $ -8\!^{+ 17}_{- 19}\!^{+  8}_{- 30}\!^{+ 20}_{-  3}$ & $< 0.4\times 10^{-3}$ & $< 9.2\times 10^{-4}$ & $< 4.5$                                                                                                  \\
                                        $K^\ast(892)^-\pi^+ \times B(K^\ast(892)^-\to{\overline K}^0\pi^-)$              & $-0.002\!\pm\!0.012\!^{+0.008}_{-0.003}\!^{+0.002}_{-0.002}$ & $ -3\!^{+ 16}_{- 18}\!^{+  9}_{- 30}\!^{+ 21}_{-  3}$ & $< 2.1\times 10^{-3}$ & $< 3.5\times 10^{-4}$ & $< 0.3$                                                                                                  \\
                                        $\overline K^0 f_0(980) \times B(f_0(980) \to \pi^+ \pi^-)$                     & $-0.04\!\pm\!0.06\!^{+0.13}_{-0.04}\!^{+0.00}_{-0.04}$ & $  9\!^{+ 16}_{- 17}\!^{+ 10}_{- 29}\!^{+ 20}_{-  3}$ & $< 4.2\times 10^{-3}$ & $< 6.8\times 10^{-4}$ & $<10.4$                                                                                                  \\
                                        $\overline K^0 f_2(1270) \times B(f_2(1270) \to \pi^+ \pi^-)$                   & $0.16\!^{+0.28}_{-0.27}\!^{+0.15}_{-0.37}\!^{+0.11}_{-0.18}$ & $ 22\!^{+ 19}_{- 20}\!^{+ 12}_{- 32}\!^{+ 20}_{-  2}$ & $< 4.4\times 10^{-3}$ & $<13.5\times 10^{-4}$ & $<150$                                                                                                  \\
                                        $\overline K^0 f_0(1370) \times B(f_0(1370) \to \pi^+ \pi^-)$                   & $0.08\!^{+0.06}_{-0.05}\!^{+0.01}_{-0.11}\!^{+0.06}_{-0.03}$ & $  8\!^{+ 15}_{- 17}\!^{+ 10}_{- 28}\!^{+ 20}_{-  4}$ & $<22\times 10^{-3}$ & $<25.5\times 10^{-4}$ & $<21$                                                                                                  \\
                                        $K^\ast_0(1430)^-\pi^+ \times
                                        B(K^\ast_0(1430)^-\to\overline
                                        K^0\pi^-)$         & $-0.02\!\pm\!0.06\!^{+0.04}_{-0.02}\!^{+0.00}_{-0.01}$ & $ -3\!^{+ 17}_{- 19}\!^{+ 13}_{- 36}\!^{+ 23}_{-  2}$ & $< 9.1\times 10^{-3}$ & $< 9.0\times 10^{-4}$ & $<14$                                                                                                  \\
                                        $K^\ast_2(1430)^-\pi^+ \times
                                        B(K^\ast_2(1430)^-\to\overline
                                        K^0\pi^-)$         & $-0.05\!\pm\!0.12\!^{+0.04}_{-0.14}\!^{+0.04}_{-0.00}$ & $  3\!^{+ 17}_{- 18}\!^{+ 10}_{- 31}\!^{+ 21}_{-  2}$ & $< 2.2\times 10^{-3}$ & $< 6.5\times 10^{-4}$ & $<22$                                                                                                  \\
                                        $K^\ast(1680)^- \pi^+ \times
                                        B(K^\ast(1680)^- \to \overline
                                        K^0 \pi^-)$         & $-0.20\!^{+0.28}_{-0.27}\!^{+0.05}_{-0.22}\!^{+0.02}_{-0.01}$ & $ -3\!^{+ 19}_{- 20}\!^{+ 20}_{- 25}\!^{+ 27}_{-  2}$ & $<19\times 10^{-3}$ & $<28.4\times 10^{-4}$ & $<92$                                                                                                  \\

\end{tabular} 
\end{ruledtabular}

\caption{
$\DZ$ and $\DZB$ Samples. Errors are
statistical, experimental systematic and modeling systematic,
respectively. 
The fit fraction is computed from Eqs.~\ref{eqn:ddpintD0} and~\ref{eqn:ddpintD0B}
following the prescription described in the text. The Fit Fraction Asymmetry is computed as the difference between the $\DZ$ and $\DZB$ Fit Fractions divided by the sum.}
\label{tbl:fit3}
\begin{ruledtabular}
\begin{tabular}{l|ccc}
Component    &
$\DZ$ Fit Fraction (\%) & $\DZB$ Fit Fraction (\%) & Fit Fraction Asymmetry (\%)\\ \hline
                                        $K^\ast(892)^+\pi^- \times B(K^\ast(892)^+\to K^0 \pi^+)$                        & $ 0.27\!\pm\! 0.20\!^{+ 0.39}_{- 0.11}\!^{+ 0.10}_{- 0.03}$ & $ 0.41\!\pm\! 0.21\!^{+ 0.39}_{- 0.00}\!^{+ 0.09}_{-   0.05}$ & $-21\!\pm\!  42\!^{+ 17}_{- 28}\!^{+ 22}_{-  4}$                                                                                                     \\
                                        $\overline K^0 \rho^0$                                                          & $27.5\!\pm\!1.6\!^{+ 1.4}_{- 0.8}\!^{+ 0.4}_{- 2.7}$ & $25.9\!\pm\!1.5\!^{+ 0.5}_{- 0.8}\!^{+ 0.7}_{-   2.7}$ & $  3.1\!\pm\! 3.8\!^{+  2.7}_{-  1.8}\!^{+  0.4}_{-  1.2}$                                                                                                     \\
                                        $\overline K^0 \omega \times B(\omega \to \pi^+\pi^-)$                          & $ 0.61\!\pm\! 0.24\!^{+ 0.32}_{- 0.09}\!^{+ 0.15}_{- 0.09}$ & $ 1.03\!\pm\! 0.31\!^{+ 0.10}_{- 0.21}\!^{+ 0.19}_{-   0.07}$ & $-26\!\pm\! 24\!^{+ 22}_{-  2}\!^{+  2}_{-  4}$                                                                                                     \\
                                        $K^\ast(892)^-\pi^+ \times B(K^\ast(892)^-\to{\overline K}^0\pi^-)$              & $68.0\!\pm\! 1.8\!^{+ 1.1}_{- 3.2}\!^{+ 4.6}_{- 3.6}$ & $64.7\!\pm\! 1.7\!^{+ 0.3}_{- 2.2}\!^{+ 0.8}_{-   3.0}$ & $  2.5\!\pm\!  1.9\!^{+  1.5}_{-  0.7}\!^{+  2.9}_{-  0.3}$                                                                                                     \\
                                        $\overline K^0 f_0(980) \times B(f_0(980) \to \pi^+ \pi^-)$                     & $ 4.0\!\pm\! 0.8\!^{+ 1.1}_{- 0.1}\!^{+ 0.2}_{- 0.4}$ & $ 4.4\!\pm\! 0.7\!^{+ 1.1}_{- 1.3}\!\pm\!0.2$ & $ -4.7\!\pm\!  11.0\!^{+ 24.9}_{-  7.4}\!^{+ 0.3}_{-  4.8}$                                                                                                     \\
                                        $\overline K^0 f_2(1270) \times B(f_2(1270) \to \pi^+ \pi^-)$                   & $ 0.49\!\pm\! 0.41\!^{+ 0.61}_{- 0.24}\!^{+ 0.14}_{- 0.29}$ & $ 0.24\!\pm\! 0.23\!^{+ 0.71}_{- 0.03}\!^{+ 0.06}_{-   0.09}$ & $ 34\!\pm\!  51\!^{+ 25}_{- 71}\!^{+ 21}_{- 34}$                                                                                                     \\
                                        $\overline K^0 f_0(1370) \times B(f_0(1370) \to \pi^+ \pi^-)$                   & $11.7\!\pm\! 1.9\!^{+ 2.3}_{- 4.2}\!^{+ 0.7}_{- 2.3}$ & $ 8.2\!\pm\! 1.7\!^{+ 2.5}_{- 0.5}\!^{+ 1.5}_{-   3.1}$ & $ 18\!\pm\!  10\!^{+  2}_{- 21}\!^{+ 13}_{-  6}$                                                                                                     \\
                                        $K^\ast_0(1430)^-\pi^+ \times
                                        B(K^\ast_0(1430)^-\to\overline
                                        K^0\pi^-)$         & $ 7.1\!\pm\!1.1\!^{+ 0.9}_{- 0.7}\!^{+ 1.5}_{- 0.6}$ & $ 7.2\!\pm\!1.1\!^{+ 0.4}_{- 1.4}\!^{+ 1.2}_{-   0.7}$ & $ -0.2\!\pm\! 11.3\!^{+  8.6}_{-  4.9}\!^{+  1.9}_{-  1.0}$                                                                                                     \\
                                        $K^\ast_2(1430)^-\pi^+ \times
                                        B(K^\ast_2(1430)^-\to\overline
                                        K^0\pi^-)$         & $ 1.0\!\pm\! 0.4\!\pm\!0.4\!^{+ 0.5}_{- 0.1}$ & $ 1.1\!\pm\! 0.3\!^{+ 0.5}_{- 0.1}\!^{+ 0.4}_{-   0.3}$ & $ -7\!\pm\!  25\!^{+  8}_{- 26}\!^{+  10}_{-  1}$                                                                                                     \\
                                        $K^\ast(1680)^- \pi^+ \times
                                        B(K^\ast(1680)^- \to \overline
                                        K^0 \pi^-)$         & $ 1.5\!\pm\! 0.6\!^{+ 0.6}_{- 1.0}\!^{+ 0.3}_{- 0.8}$ & $ 3.2\!\pm\!  0.8\!^{+ 0.7}_{- 0.2}\!^{+ 0.3}_{-   1.7}$ & $-36\!\pm\!  19\!^{+  9}_{- 35}\!^{+  5}_{-  1}$                                                                                                     \\

\end{tabular} 
\end{ruledtabular}

\end{table*} 

We use the full covariance matrix~\cite{covar} from the fits to determine
the errors on fit fractions and the $CP$ violating interference fractions
so that the assigned errors will properly
include the correlated components of the errors on the amplitudes and
phases. After each fit, the covariance matrix and final parameter
values are used to generate 500 sample parameter sets.  For each set,
the fit fractions are calculated and recorded in histograms. The
statistical error on the fit fractions is then extracted from the histograms.
In Table \ref{tbl:fit1}, we report the results for the $CP$ conserving fit
fractions, and the 95\% upper limit for $CP$ violating contributions
are given in Table~\ref{tbl:fit2}. The fit fractions for the $\DZ$ and
$\DZB$ samples are given in Table~\ref{tbl:fit3}.
An alternative measure of the rate of $CP$ violation in a given submode
is the asymmetry between the $\DZ$ and $\DZB$ fit fractions, 
which are also given in Table III.  
The ``Fit Fraction Asymmetry'' is similar in sensitivity to $A_{CP}$ 
defined by Eq.~\ref{eqn:acpj}.

The common evaluation of the integrated $CP$ asymmetry between normalized 
amplitudes squared across the Dalitz plot is sensitive to an asymmetry
in shape between the $\DZ$ and $\DZB$ samples and is defined
as
\begin{equation}
{\cal A}_{CP}\! =\! \int { |{\cal M}|^2 - |{\cal \overline M}|^2 \over{
|{\cal M}|^2 + |{\cal \overline M}|^2 }}
dm^2_{RS}dm^2_{\pi\pi} / \int {dm^2_{RS}dm^2_{\pi\pi}}.
\label{eqn:acpold}
\end{equation}
We obtain 
${\cal A}_{CP}\!=\!-0.009\! \pm\! 0.021 ^{+0.010}_{-0.043}\!^{+0.013}_{-0.037}$
where the errors are
statistical, experimental systematic and modeling systematic,
respectively.

We consider systematic uncertainties from experimental sources and
from the decay model separately.
Our general procedure is to change some aspect of our fit and
interpret the change in the values of the amplitudes, phases and fit
fractions in the non-standard fit relative to our nominal fit as an
estimate of the systematic uncertainty. The impact of systematic uncertainties
on the upper limit of the interference fraction, $CP$ violating fit fraction and $A_{CP}$ reported in Table~\ref{tbl:fit2} are estimated by recomputing the 
statistical error on these quantities with the covariance matrix of the 
non-standard fits using procedure described above.
Contributions to the experimental systematic uncertainties arise
from our 
model of the background, the efficiency, and biases due to experimental
resolution.
The background is modeled with a two dimensional cubic polynomial and
the covariance matrix of the polynomial coefficents, both determined
from a sideband.  Our nominal fit fixes the coefficients of the background
polynomial, and to estimate the systematic uncertainty on 
this background shape we perform a fit with the coefficients allowed
to float constrained by the covariance matrix.
Similarly we perform a fit with a uniform efficiency, rather than
the nearly uniform efficiency determined from the simulation, as estimates
of the systematic uncertainty due to the efficiency. We also perform
fits where the background normalization and the efficiency of the 
$\DZ$ and $\DZB$ samples
are determined separately.
We compute the overall normalization by evaluating the integrals in
Eqs.~\ref{eqn:normm} and \ref{eqn:normbkgd}
using Gaussian quadrature to interpolate 
between points on a finite grid across the Dalitz plot.
To study the effect of the finite resolution our experiment has on the
variables in the Dalitz plots we vary the granularity of the grid 
used to compute the overall normalization. 

We change selection criteria in the
analysis to test whether our simulation properly models the efficiency.
We introduce a track momentum cut of 350~MeV$/c$ to avoid the
difficulty of modeling our low momentum tracking efficiency. We expand
the signal region from three to six standard deviations in $Q$, $M$, and
$m_{K^0_S}$, and increase the $p_{D^\ast}$ cut from 2.0 to 3.0 GeV$/c$.
These variations to the nominal fit are the largest contribution to
our experimental systematic errors.

	Contributions to the theoretical systematic uncertainties
arise from our choices
for the decay model for $\DZ \to K_S^0\pi^+\pi^-$.  
The standard value for the radius parameter~\cite{blatweis} for
the intermediate resonances and for the $\DZ$ is
$0.3\ {\rm fm}$ and $1\ {\rm fm}$, respectively.
We vary the radius parameter between zero and twice
the standard value.
Additionally, we allow the masses and widths for the intermediate resonances to
vary within their known errors~\cite{pdg,Aitala:2000xt,Kirk:2000ay}.  

We consider the uncertainty
arising from our choice of resonances included in the fit.
We compared the result of our nominal fit to a series of fits where each of
the resonances, $\sigma$ or $f_0(600)$, $f_0(1500)$, $f_0(1710)$, $\rho(1450)$, 
$\rho(1700)$ were included one at a time. We also
considered a fit including both the $f_0(600)$ and $f_0(1500)$
resonances. These variations to the nominal fit result in highly
asymmetric variation in fit parameters and are the largest 
contribution to our modeling systematic error.

We take the 
maximum variation of the
amplitudes, phases and fit fractions 
from the nominal result compared to the results in
this series of fits as a measure of the experimental systematic 
and modeling systematic uncertainty.

In conclusion, we have analyzed the resonant substructure of 
the decay $\DZ$ or $\DZB \to K_S^0\pi^+\pi^-$
using the Dalitz-plot analysis technique and 
searched for $CP$ violation in the amplitudes and phases 
of the ten clearly observed intermediate 
resonances. 
Our results, shown in Table~\ref{tbl:fit2} and 
Table~\ref{tbl:fit3}, are consistent with the absence of $CP$
violation. 
We find the $CP$ asymmetry in
the fit fractions for each decay channel to be in the range $<(3.5
{\rm \,to\, } 28.4) \times 10^{-4}$ at the 95\% confidence level.
We find the $CP$ asymmetry in
the interference fractions for each decay channel to be in the range
$<(0.4 {\rm \,to\, }22) \times 10^{-3}$ at the 95\% confidence level.
We find the ratio of the $CP$ violating to $CP$ conserving rate for
each decay channel, to be in the range
$<(0.3 {\rm \,to\, } 150)\%$ at the 95\% confidence level.
We find $A_{CP}$ which is the asymmetry between 
normalized squared amplitudes
integrated over the entire Dalitz
plot to be $-0.009\! \pm\! 0.021 ^{+0.010}_{-0.043}\!^{+0.013}_{-0.037}$.

\section*{Acknowledgment}

We thank Eugene Golowich and Jon Rosner for valuable discussions.
We gratefully acknowledge the effort of the CESR staff in providing us with
excellent luminosity and running conditions.
This work was supported by 
the National Science Foundation and
the U.S. Department of Energy.

\end{document}